\documentclass[a4paper,11pt]{article}

\usepackage[margin=25mm]{geometry}
\usepackage{amsmath,amssymb}
\usepackage{upgreek}
\usepackage{tcolorbox}
\usepackage{tikz}
\usetikzlibrary{shapes.geometric,arrows.meta}
\usepackage{authblk}
\usepackage{cite}
\usepackage[colorlinks=true,linkcolor=blue,citecolor=blue,urlcolor=blue]{hyperref}
\usepackage[capitalize]{cleveref}
\hypersetup{pdftitle={Emergent Symmetries in Ensemble Averages},pdfauthor={Masahito Yamazaki}}
\usepackage{fancyhdr}
\usepackage{url}

\numberwithin{equation}{section}

\fancypagestyle{titlepage}{
  \fancyhf{}
  \rhead{\small RIKEN-iTHEMS-Report-26}
  
}

\newcommand{\bbZ}{\mathbb{Z}}
\newcommand{\bbR}{\mathbb{R}}
\newcommand{\calM}{\mathcal{M}}
\newcommand{\calO}{\mathcal{O}}
\newcommand{\calD}{\mathcal{D}}
\newcommand{\calH}{\mathcal{H}}
\newcommand{\Tr}{\mathrm{Tr}}
\newcommand{\PSL}{\mathrm{PSL}}
\newcommand{\SL}{\mathrm{SL}}
\newcommand{\GL}{\mathrm{GL}}
\newcommand{\Vol}{\mathrm{Vol}}
\newcommand{\Aut}{\mathrm{Aut}}

\title{\vspace{3cm}Emergent Symmetries in Ensemble Averages}

\author[1,2,3,4]{Masahito Yamazaki\thanks{\href{mailto:masahito.yamazaki@ipmu.jp}{masahito.yamazaki@ipmu.jp}}}
\affil[1]{\small Department of Physics, University of Tokyo, Tokyo 113-0033, Japan}
\affil[2]{\small Kavli Institute for Physics and Mathematics of the Universe (Kavli IPMU),\protect\newline UTIAS, University of Tokyo, Chiba 277-8583, Japan}
\affil[3]{\small Trans-Scale Quantum Science Institute, University of Tokyo, Tokyo 113-0033, Japan}
\affil[4]{\small Center for Interdisciplinary Theoretical and Mathematical Sciences (iTHEMS), \protect\newline RIKEN, Saitama 351-0198, Japan}
\date{}

\begin{document}
\maketitle
\thispagestyle{titlepage}

\vspace{1.5cm}
\begin{abstract} 
  It is widely believed that there are no exact global symmetries in quantum gravity. We review recent works showing that global symmetries can nevertheless emerge after ensemble averages of theories, as encountered in holography. As a precision case study, we discuss the ensemble average of two-dimensional Narain-type conformal field theories associated with an even quadratic form of general signature, which is computed by the Siegel--Weil formula and interpreted as a sum over geometries in a three-dimensional Abelian Chern-Simons theory. Global symmetries of the bulk anyons can emerge after the average, as T-dualities relating different theories in the ensemble are ``folded'' into symmetries of a single theory (duality origami). We discuss the relation to the Swampland program, including the emergence of global symmetries at infinite distance. Finally, we recast strong-to-weak spontaneous symmetry breaking (SWSSB) of mixed states as an emergence of symmetries in ensemble averages; in holography, wormholes connecting the two sides of a thermofield-double construction provide its bulk realization. This motivates a new Swampland conjecture.
\end{abstract}
 
\clearpage
 
\tableofcontents

\section{Introduction}\label{sec:intro}

Symmetry has been one of the guiding principles of physics. It is therefore natural to ask what role symmetries play in theories of quantum gravity.

The canonical answer to this question is that there are no exact global symmetries in quantum gravity \cite{Misner:1957mt,Banks:1988yz,Kamionkowski:1992mf,Kallosh:1995hi,Banks:2010zn,Harlow:2018jwu,Harlow:2018tng,Hsin:2020mfa}. This is one of the best-known conjectures in the Swampland program \cite{Vafa:2005ui,Ooguri:2006in}, which posits the existence of non-trivial consistency conditions for low-energy effective field theories to be embedded into ultraviolet-complete theories of quantum gravity. In the context of the AdS/CFT correspondence \cite{Maldacena:1997re}, global symmetries of the boundary theory are realized as gauge symmetries in the bulk, and the absence of global symmetries in the bulk was argued from the holographic viewpoint \cite{Harlow:2018jwu,Harlow:2018tng}.

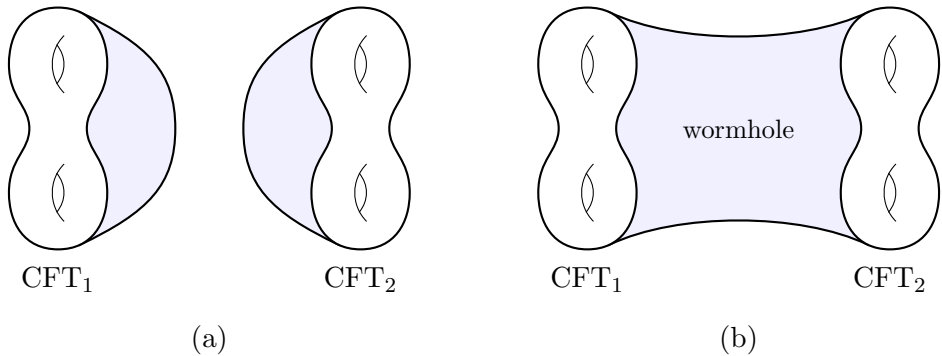
\begin{figure}[htbp] 
  \centering
  \begin{tikzpicture}[
      genustwo/.pic={
        \draw[thick,fill=white] (0,1.6) .. controls (0.45,1.6) and (0.65,1.25) .. (0.65,0.85)
          .. controls (0.65,0.4) and (0.38,0.3) .. (0.38,0)
          .. controls (0.38,-0.3) and (0.65,-0.4) .. (0.65,-0.85)
          .. controls (0.65,-1.25) and (0.45,-1.6) .. (0,-1.6)
          .. controls (-0.45,-1.6) and (-0.65,-1.25) .. (-0.65,-0.85)
          .. controls (-0.65,-0.4) and (-0.38,-0.3) .. (-0.38,0)
          .. controls (-0.38,0.3) and (-0.65,0.4) .. (-0.65,0.85)
          .. controls (-0.65,1.25) and (-0.45,1.6) .. (0,1.6) -- cycle;
        \foreach \y in {0.85,-0.85}{
          \draw (0.08,\y+0.38) to[bend right=45] (0.08,\y-0.38);
          \draw (-0.02,\y+0.26) to[bend left=40] (-0.02,\y-0.26);
        }
      }]
    \draw[thick,fill=blue!6] (0.291,1.539) .. controls (1.19,1.11) and (1.55,0.75) .. (1.55,0)
      .. controls (1.55,-0.75) and (1.19,-1.11) .. (0.291,-1.539) -- cycle;
    \draw[thick,fill=blue!6] (3.709,1.539) .. controls (2.81,1.11) and (2.45,0.75) .. (2.45,0)
      .. controls (2.45,-0.75) and (2.81,-1.11) .. (3.709,-1.539) -- cycle;
    \pic at (0,0) {genustwo};
    \pic at (4,0) {genustwo};
    \node at (0,-2) {CFT$_1$};
    \node at (4,-2) {CFT$_2$};
    \node at (2,-2.8) {(a)};
    \begin{scope}[xshift=7cm]
      \draw[thick,fill=blue!6] (0.291,1.539) .. controls (1.19,1.11) and (2.81,1.11) .. (3.709,1.539)
        -- (3.709,-1.539) .. controls (2.81,-1.11) and (1.19,-1.11) .. (0.291,-1.539) -- cycle;
      \pic at (0,0) {genustwo};
      \pic at (4,0) {genustwo};
      \node at (2,0) {\small wormhole};
      \node at (0,-2) {CFT$_1$};
      \node at (4,-2) {CFT$_2$};
      \node at (2,-2.8) {(b)};
    \end{scope}
  \end{tikzpicture}
  \caption{The factorization puzzle \cite{Witten:1999xp,Maldacena:2004rf}. The two boundary CFTs live on (here genus-two) Riemann surfaces, drawn in white, and the shaded regions represent the bulk, whose boundaries are these surfaces. (a) Disconnected bulk geometries, each filling in one of the boundaries, reproduce the factorized answer $Z_{\mathrm{CFT}_1} \cdot Z_{\mathrm{CFT}_2}$. (b) The gravitational path integral also includes wormhole geometries connecting the two boundaries, whose contributions do not factorize. The puzzle can be resolved if the bulk computes the ensemble average on the boundary, for which $\langle Z_{\mathrm{CFT},1}\rangle \langle Z_{\mathrm{CFT},2} \rangle \ne \langle Z_{\mathrm{CFT},1} \cdot Z_{\mathrm{CFT},2} \rangle$.}
  \label{fig:factorization}
\end{figure}

In this article we discuss a twist to this story, which arises from the concept of \emph{ensemble averages}. Recall that the standard holographic dualities postulate a one-to-one correspondence between a bulk theory of quantum gravity and a boundary quantum field theory. This leads to a puzzle, known as the factorization puzzle \cite{Witten:1999xp,Maldacena:2004rf}. Suppose that we have two copies of the boundary conformal field theories (CFTs), CFT$_1$ and CFT$_2$. Since the two theories are decoupled, the total partition function should factorize: $Z_{\mathrm{tot}}=Z_{\mathrm{CFT}_1} \cdot Z_{\mathrm{CFT}_2}$. On the gravity side, this factorization is reproduced by the disconnected geometries, where each boundary is filled in separately (\cref{fig:factorization}(a)). However, the gravitational path integral also includes contributions from wormhole geometries connecting the two boundaries (\cref{fig:factorization}(b)), and such contributions do not factorize. One possible resolution of the puzzle is to regard the boundary theory as an ensemble of theories, and to interpret the gravitational path integral as computing the ensemble average $\langle - \rangle$ of the boundary quantities. Since in general
\begin{align}
  \langle Z_{\mathrm{CFT},1}\rangle \langle Z_{\mathrm{CFT},2} \rangle \ne \langle Z_{\mathrm{CFT},1} \cdot Z_{\mathrm{CFT},2} \rangle \;,
\end{align}
there is no contradiction with the existence of wormholes. Such ensemble averages have indeed been identified for two-dimensional Jackiw--Teitelboim gravity \cite{Saad:2019lba,Stanford:2019vob} and for ensembles of Narain CFTs \cite{Afkhami-Jeddi:2020ezh,Maloney:2020nni} (see also \cite{Cotler:2020ugk,Cotler:2020hgz,Perez:2020klz,Benjamin:2021wzr,Ashwinkumar:2021kav,Dong:2021wot,Collier:2021rsn,Ashwinkumar:2023eit} for further developments). The relation between ensemble averages, wormholes, and baby universes goes back to the early days \cite{Coleman:1988cy,Giddings:1988cx,Giddings:1987cg}; see \cite{Marolf:2020xie,McNamara:2020uza,Heckman:2021vzx} for more recent discussions.

We should emphasize that the ensembles here are ensembles of \emph{theories}. While we routinely consider ensembles inside a single theory, e.g.\ in statistical mechanics, here we consider averages of the form
\begin{align}
  \langle \calO \rangle = \int_{\calM} [dm]\, \calO(m) \;, 
  \label{eq:ensemble_intro}
\end{align}
where $\calM$ is the moduli space of theories, $[dm]$ is a normalized measure on $\calM$, and $m$ denotes the parameters specifying a theory inside the ensemble.

As we will review in this article, ensemble averages provide an interesting loophole in the argument against global symmetries in quantum gravity: a symmetry absent in a generic theory in the ensemble can \emph{emerge} after the ensemble average \cite{Ashwinkumar:2023jtz} (see also \cite{Antinucci:2023uzq,Torres:2025jcb}). This review is based on four papers by the author and collaborators \cite{Ashwinkumar:2021kav,Ashwinkumar:2023jtz,Ashwinkumar:2023eit,Kawamoto:2026jdz}, and the discussion is organized as follows.

In \cref{sec:case_study} we discuss a precision case study, namely the ensemble average of the generalized Narain theories associated with a general even quadratic form \cite{Ashwinkumar:2021kav,Ashwinkumar:2023eit}. In this case the ensemble average can be computed exactly, and Abelian Chern-Simons theory describes the perturbative contribution on each bulk geometry. In \cref{sec:emergent} we discuss how some global symmetries of the bulk anyons emerge after the ensemble average, and formulate the general mechanism (``duality origami'') \cite{Ashwinkumar:2023jtz}. In \cref{sec:swampland} we come back to the Swampland program and discuss the relation between the emergence of global symmetries in ensemble averages and that in the infinite distance limit \cite{Ashwinkumar:2023jtz}. In \cref{sec:SWSSB} we discuss a more general class of ensembles, namely density matrices, and show that strong-to-weak spontaneous symmetry breaking (SWSSB) can be recast as an emergence of symmetries in ensemble averages \cite{Kawamoto:2026jdz}. We also discuss the realization of SWSSB in holography, and propose a new Swampland conjecture. We conclude in \cref{sec:summary} with a summary and future directions.

\section{Precision Case Study: Generalized Narain Theories}\label{sec:case_study}

In this section we discuss the ensemble average of the generalized Narain theories\footnote{We can generalize the discussion to orbifolds of the generalized Narain theories, where the moduli space is projected to a smaller subspace, and to theories with odd quadratic forms, which depend on the choice of the spin structure \cite{Ashwinkumar:2021kav,Ashwinkumar:2023eit}. We will not discuss these generalizations in this article.} and their holographic duals, following \cite{Ashwinkumar:2021kav,Ashwinkumar:2023eit}. This generalizes the previous analysis \cite{Afkhami-Jeddi:2020ezh,Maloney:2020nni} of standard Narain CFTs \cite{Narain:1985jj,Narain:1986am}, defined by even self-dual lattices, to arbitrary even lattices of general signature.

\subsection{Data}\label{subsec:data}

The discrete data specifying a generalized Narain theory is an even integral quadratic form $Q$ of rank $p+q$ and signature $(p,q)$:
\begin{align}
  Q(\ell) = \sum_{i,j=1}^{p+q} Q_{ij}\, \ell^i \ell^j \in 2\bbZ \;, \qquad \ell \in \Lambda := \bbZ^{p+q} \;.
  \label{eq:Q}
\end{align}
The quadratic form defines an even integral lattice $\Lambda$, with the bilinear form
\begin{align}
  Q(\ell, \ell') := \frac{Q(\ell+\ell')-Q(\ell)-Q(\ell')}{2} = \sum_{i,j=1}^{p+q} Q_{ij}\, \ell^i \ell'^j \in \bbZ \;.
\end{align}
We do not require $\Lambda$ to be self-dual. We define the dual lattice $\Lambda^*:=\{ x\in \bbR^{p+q} \,|\, Q(x,\ell)\in \bbZ \; (\forall \ell \in \Lambda) \}$, which contains $\Lambda$ as a sublattice. Here and below we extend the bilinear form to $\bbR^{p+q}$ and write $Q(x):=Q(x,x)$ for $x\in\bbR^{p+q}$, which agrees with \eqref{eq:Q} on $\Lambda$. We also define the discriminant group
\begin{align}
  \calD := \Lambda^*/\Lambda \;, \qquad |\calD| = |\det Q| \;.
  \label{eq:D}
\end{align}
The lattice $\Lambda$ is self-dual if and only if $\calD$ is trivial.

The continuous data, i.e.\ a point $m$ of the moduli space of the CFT, is specified by a decomposition of the quadratic form $Q$ into left- and right-moving parts, $Q(\ell)=p_L^2 - p_R^2$. Equivalently, this is specified by a positive definite quadratic form, the Hamiltonian
\begin{align}
  H(\ell)=p_L^2 + p_R^2 \;,
\end{align}
which is moduli dependent and satisfies $H Q^{-1} H = Q$. The moduli space of the theory is the double coset \cite{Ashwinkumar:2021kav}
\begin{align}
  \calM_Q = O_Q(p,q;\bbZ) \backslash O(p,q;\bbR) / (O(p;\bbR)\times O(q;\bbR)) \;, \qquad \dim_{\bbR} \calM_Q = pq \;,
  \label{eq:M_Q}
\end{align}
where
\begin{align}
  O_Q(p,q;\bbZ) := \{ \Sigma \in \GL(p+q,\bbZ) \,|\, \Sigma^T Q \Sigma = Q \}
  \label{eq:T-duality}
\end{align}
is the T-duality group of the theory.

As the simplest example, consider the $S^1$-compactification of the free boson, where $p=q=1$. The left- and right-moving momenta are given by
\begin{align}
  p_L = \frac{n}{2R} + w R \;, \qquad
  p_R = \frac{n}{2R} - w R \;, \qquad (n, w\in \bbZ) \;,
\end{align}
where $n$ and $w$ are momentum and winding, and $R$ is the radius of the circle. We then have
\begin{align}
  Q = p_L^2 - p_R^2 = 2 n w \in 2\bbZ \;, \qquad
  H = p_L^2 + p_R^2 = \frac{n^2}{2R^2} + 2 w^2 R^2 \;.
\end{align}
Note that $Q$ is independent of the modulus $R$, while $H$ depends on $R$. The moduli space is the half line $R\ge 1/\sqrt{2}$ (in our normalization), where the T-duality $R\to 1/(2R)$ exchanges $n$ and $w$.

\subsection{Theta Functions}\label{subsec:theta}

The torus partition functions of the generalized Narain theory are labeled by elements $\alpha\in \calD$ of the discriminant group:
\begin{align}
  Z_{Q,\alpha}(\tau,\bar{\tau};m) = \frac{\vartheta_{Q,\alpha}(\tau,\bar{\tau};m)}{\eta(\tau)^p\, \overline{\eta(\tau)}^q} \;,
\end{align}
where $\tau=\tau_1+i \tau_2$ is the modulus of the spacetime torus, $\eta(\tau)$ is the Dedekind eta function, and $\vartheta_{Q,\alpha}$ is the Siegel--Narain theta function
\begin{align}
  \vartheta_{Q,\alpha}(\tau,\bar{\tau};m) := \sum_{\ell\in \Lambda} e^{\pi i \tau_1 Q(\ell+\alpha) - \pi \tau_2 H(\ell+\alpha)} \;.
  \label{eq:theta}
\end{align}
The dependence on the CFT moduli $m$ is encoded in $H$.

Under the generators of $\SL(2,\bbZ)$, the theta functions transform as \cite{Ashwinkumar:2021kav}
\begin{align}
  \begin{split}
    T: &\quad \vartheta_{Q,\alpha}(\tau+1;m) = e^{\pi i Q(\alpha)}\, \vartheta_{Q,\alpha}(\tau;m) \;, \\
    S: &\quad \vartheta_{Q,\alpha}\left(-\frac{1}{\tau};m\right) = \frac{e^{-\pi i \sigma/4}}{\sqrt{|\det Q|}}\, \tau^{\frac{p}{2}} \bar{\tau}^{\frac{q}{2}} \sum_{\beta\in \calD} e^{-2\pi i Q(\alpha,\beta)}\, \vartheta_{Q,\beta}(\tau;m) \;,
  \end{split}
  \label{eq:theta_modular}
\end{align}
where $\sigma:=p-q$. The partition functions therefore form a vector-valued modular form, and are not modular invariant unless $\Lambda$ is self-dual and $\sigma\equiv 0\pmod{24}$. The two-dimensional theory is then naturally regarded as a relative theory with respect to the three-dimensional bulk. Note also that the transformation matrices in \eqref{eq:theta_modular} are independent of the moduli. This will play a crucial role below.

\subsection{Ensemble Average and the Siegel--Weil Formula}\label{subsec:Siegel-Weil}

Let us next consider the ensemble average of the theta function over the moduli space:
\begin{align}
  \langle \vartheta_{Q,\alpha} \rangle (\tau,\bar{\tau}) := \frac{1}{\Vol(\calM_{Q,\alpha})} \int_{\calM_{Q,\alpha}} d\nu(m)\, \vartheta_{Q,\alpha}(\tau,\bar{\tau};m) \;,
  \label{eq:average}
\end{align}
where $d\nu(m)$ is the unnormalized invariant measure on the moduli space, induced by the Zamolodchikov metric, and $\Vol(\calM_{Q,\alpha}):=\int_{\calM_{Q,\alpha}}d\nu(m)$. Since the T-duality group \eqref{eq:T-duality} in general acts non-trivially on the labels $\alpha\in \calD$, the region of integration $\calM_{Q,\alpha}$ is the quotient by the subgroup of $O_Q(p,q;\bbZ)$ preserving $\alpha$ \cite{Ashwinkumar:2023jtz,Ashwinkumar:2023eit}. We impose $p+q>4$ for the convergence of the integral.

A remarkable theorem by Siegel \cite{Siegel:1951}, later generalized by Weil \cite{Weil:1964,Weil:1965} and therefore known as the Siegel--Weil formula, states that for $pq\neq 0$ the ensemble average is given by
\begin{align}
  \langle \vartheta_{Q,\alpha} \rangle (\tau,\bar{\tau}) = E_{Q,\alpha}(\tau,\bar{\tau}) \;,
  \label{eq:Siegel-Weil}
\end{align}
where $E_{Q,\alpha}$ is the Siegel-Eisenstein series associated with the quadratic form $Q$:
\begin{align}
  E_{Q,\alpha}(\tau,\bar{\tau}) := \delta_{\alpha\in \Lambda} + \sum_{\substack{(c,d)=1\\ c>0}} \frac{\gamma_{Q,\alpha}(c,d)}{(c\tau+d)^{\frac{p}{2}}(c\bar{\tau}+d)^{\frac{q}{2}}} \;.
  \label{eq:Eisenstein}
\end{align}
Here $\delta_{\alpha\in \Lambda}=1$ for $\alpha\in \Lambda$ and $0$ otherwise, and the sum is over pairs of coprime integers $(c,d)$ with $c>0$ (equivalently, over rational numbers $d/c$). The factor $\gamma_{Q,\alpha}(c,d)$ is a version of the quadratic Gauss sum:
\begin{align}
  \gamma_{Q,\alpha}(c,d) := \frac{e^{\frac{\pi i \sigma}{4}}}{\sqrt{|\det Q|}}\, c^{-\frac{p+q}{2}} \sum_{\ell\in \Lambda/c\Lambda} \exp\left[ -\pi i \frac{d}{c}\, Q(\ell+\alpha) \right] \;.
  \label{eq:gamma}
\end{align}
The Eisenstein series is a non-holomorphic modular form which transforms in the same way as the theta functions \eqref{eq:theta_modular}, as it should. For the standard Narain case with even self-dual lattice $Q=\mathrm{II}_{p,p}$ we have $\gamma_{Q}(c,d)=1$, and \eqref{eq:Eisenstein} reduces to the non-holomorphic Eisenstein series $1+\sum_{(c,d)=1,c>0}|c\tau+d|^{-p}$ discussed in \cite{Afkhami-Jeddi:2020ezh,Maloney:2020nni}.

A simple proof of the Siegel--Weil formula for an even indefinite quadratic form was given in \cite{Ashwinkumar:2021kav}, along the lines of \cite{Maloney:2020nni}. The argument consists of two steps. First, both sides of \eqref{eq:Siegel-Weil} have the same behavior at the cusps of the upper half plane, namely at the images of $\tau=i\infty$ under $\PSL(2,\bbZ)$. For the theta function this follows from the modular transformations \eqref{eq:theta_modular}: near a cusp $\tau\to a/c$ we have
\begin{align}
  \vartheta_{Q,\alpha}(\tau,\bar{\tau};m) \sim \frac{\gamma_{Q,\alpha}(c,-a)}{(c\tau-a)^{\frac{p}{2}}(c\bar{\tau}-a)^{\frac{q}{2}}} \;,
  \label{eq:cusp}
\end{align}
which is independent of the moduli, and coincides with the behavior of the corresponding term in the Eisenstein series. Second, both sides of \eqref{eq:Siegel-Weil} satisfy the same differential equation
\begin{align}
  \left[ \tau_2^2 \left(\partial_{\tau_1}^2 + \partial_{\tau_2}^2\right) + \frac{p+q}{2} \tau_2 \partial_{\tau_2} + \frac{i(q-p)}{2} \tau_2 \partial_{\tau_1} \right] f(\tau,\bar{\tau}) = 0 \;.
\end{align}
One can then show that a solution to this differential equation is uniquely determined by its behavior at the cusps, since there is no square-normalizable eigenfunction of the relevant Laplacian with the required eigenvalue when $q>0$. This completes the argument. The proof uses the modular transformation laws and the differential equation for the theta functions, without explicitly integrating over the moduli space.\footnote{The same strategy can be used to derive new Siegel--Weil-type formulas, for example for orbifolds of the generalized Narain theories \cite{Ashwinkumar:2023eit}, where we obtain the ``orbifold Eisenstein series'' as a new modular form.}

\subsection{Holographic Dual after Averaging}\label{subsec:bulk}

Let us now come to the holographic interpretation of the Eisenstein series \eqref{eq:Eisenstein}. The Poincar\'e sum over coprime pairs $(c,d)$ in \eqref{eq:Eisenstein} represents a sum over $\Gamma_{\infty}\backslash \PSL(2,\bbZ)$, where $\Gamma_{\infty}\simeq \bbZ$ is generated by $T$. We interpret this as a sum over geometries in the three-dimensional bulk. Namely, the geometries are the so-called $\PSL(2,\bbZ)$ black holes $M_{(c,d)}$ \cite{Maldacena:1998bw,Dijkgraaf:2000fq}, which are solid tori whose boundary is the spacetime torus, and which differ in the choice of the cycle of the boundary torus that becomes contractible in the bulk. These geometries include thermal AdS$_3$ ($M_{(0,1)}$) and the BTZ black hole ($M_{(1,0)}$) \cite{Banados:1992wn}.

The contribution from each geometry is determined by the three-dimensional Abelian Chern-Simons theory with gauge group $U(1)^{p+q}$, whose action is determined by the quadratic form $Q$:
\begin{align}
  S_{\mathrm{CS}} = \sum_{i,j=1}^{p+q} \frac{Q_{ij}}{4\pi} \int A_i \wedge dA_j \;.
  \label{eq:S_CS}
\end{align}
The existence of the $U(1)^{p+q}$ gauge symmetry in the bulk is expected from the $U(1)^{p+q}$ global symmetry of the boundary theory, and the non-modular-invariance of the boundary partition functions is accounted for by the Chern-Simons term in the bulk. Incidentally, the Chern-Simons theory \eqref{eq:S_CS} is used in condensed matter physics to describe Abelian topological phases in two spatial dimensions, where $Q$ is known as the $K$-matrix \cite{Wen:1992uk,Lu:2012dt}. The lattices considered here thus connect the ensemble average to a broad class of such phases.

The Chern-Simons theory \eqref{eq:S_CS} is an Abelian topological quantum field theory (TQFT), whose line operators describe Abelian anyons. Wilson-line charges are integral vectors $l\in\bbZ^{p+q}$, represented in the lattice convention by $\alpha=Q^{-1}l\in\Lambda^*$. Charges differing by $Q\Lambda$ define the same anyon, so the anyon labels form $\calD=\Lambda^*/\Lambda$. Their fusion is addition in $\calD$, and the topological spin of $\alpha$ is
\begin{align}
  \uptheta(\alpha) = e^{\pi i Q(\alpha)} \;,
  \label{eq:spin}
\end{align}
and the braiding of two anyons $\alpha$ and $\beta$ generates the phase
\begin{align}
  B(\alpha,\beta) = \frac{\uptheta(\alpha+\beta)}{\uptheta(\alpha)\uptheta(\beta)} = e^{2\pi i Q(\alpha,\beta)} \;.
  \label{eq:braiding}
\end{align}
The modular $S$- and $T$-matrices of the TQFT are given by
\begin{align}
  S_{\alpha\beta} = \frac{1}{\sqrt{|\calD|}}\, e^{-2\pi i Q(\alpha,\beta)} \;, \qquad
  T_{\alpha\beta} = e^{-\frac{2\pi i\sigma}{24}}\, \uptheta(\alpha)\, \delta_{\alpha\beta} \;,
  \label{eq:ST}
\end{align}
which coincide with the modular transformations \eqref{eq:theta_modular} of the theta functions, up to the contributions from the eta functions. Indeed, the Hilbert space of the Chern-Simons theory on the boundary torus is spanned by the states $|\alpha\rangle$, $\alpha\in\calD$, which are prepared by the path integral on a solid torus with an insertion of the Wilson line $\alpha$ along the non-contractible cycle. We can thus regard the Eisenstein series $E_{Q,\alpha}(\tau,\bar{\tau})$, $\alpha\in\calD$, as the components of a wavefunction of the Chern-Simons theory on $T^2$.

This identification leads to a precise interpretation of the Gauss sum \eqref{eq:gamma}. For an element $g=\left(\begin{smallmatrix} a & b \\ c & d\end{smallmatrix}\right)$ of $\SL(2,\bbZ)$, let us denote by $U(g)$ its representation on the Hilbert space. The matrix element $\langle 0|U(g)|\alpha\rangle$ is the Chern-Simons partition function of the geometry obtained by gluing two solid tori by $g$, which is the lens space $L(c,d)$, with an insertion of the Wilson line $\alpha$. Here the lens space $L(c,d)$ is defined as the $\bbZ_c$ quotient of the three-sphere $|z_1|^2+|z_2|^2=1$ by $(z_1,z_2)\sim (e^{2\pi i/c}z_1, e^{2\pi i d/c}z_2)$. We find \cite{Ashwinkumar:2021kav}
\begin{align}
  \langle 0|U(g)|\alpha\rangle^* = \langle \alpha|U(g)^{-1}|0\rangle = e^{\frac{2\pi i\sigma\Phi(g)}{24}-\frac{\pi i \sigma}{4}}\, \gamma_{Q,\alpha}(c,d) \;, \qquad c>0 \;,
\end{align}
where the phase factor, which involves the Rademacher function $\Phi(g)$, represents the framing anomaly. The Gauss sum $\gamma_{Q,\alpha}(c,d)$ is therefore related to a lens space partition function of the Chern-Simons theory with a Wilson line insertion \cite{Witten:1988hf,Jeffrey:1992tk}. After including the boundary oscillator factors, we obtain
\begin{align}
  \langle Z_{Q,\alpha}(\tau,\bar{\tau}) \rangle = \sum_{g\in \Gamma_{\infty}\backslash \PSL(2,\bbZ)} \frac{\langle \alpha |U(g)^{-1}|0\rangle}{\eta(g\cdot \tau)^p\, \overline{\eta(g\cdot\tau)}^q} \;.
  \label{eq:Z_bulk}
\end{align}
To summarize, the ensemble average of the boundary partition function is a sum over geometries, where each geometry contributes a Chern-Simons matrix element related to a lens space partition function. The label $\alpha$ indicates a Wilson line along the non-contractible cycle of $M_{(c,d)}$. The complete bulk dual after averaging remains to be understood; the Abelian Chern-Simons theory \eqref{eq:S_CS} supplies a perturbative description on each bulk geometry. In particular, the distinction between compact and noncompact bulk gauge groups and the treatment of large gauge transformations require care \cite{Ashwinkumar:2021kav}. An alternative treatment of factorization in related Chern-Simons models gauges a one-form symmetry instead of using an ensemble sum \cite{Benini:2022hzx}.

Before the ensemble average, the holographic dual of a single theory in the ensemble is related to the Maxwell-Chern-Simons theory \cite{Gukov:2004id,Ashwinkumar:2021kav}. Schematically, the action is
\begin{align}
  S_{\mathrm{MCS}} = \frac{1}{2e^2}\int \lambda_{ij}\, dA_i \wedge * dA_j + S_{\mathrm{CS}} \;,
  \label{eq:S_MCS}
\end{align}
where $e^2$ is a dimensionful coupling and $\lambda$ is a positive definite matrix. The Maxwell term is irrelevant and the Chern-Simons term dominates in the infrared. The effect of the Maxwell term, however, still remains in the topological limit $e^2\to\infty$, since the quantization conditions of the gauge fields depend on $\lambda$, and hence on the point of the moduli space. With standard boundary conditions the wavefunctions of the Maxwell-Chern-Simons theory reproduce the theta functions $\vartheta_{Q,\alpha}(\tau,\bar{\tau};m)$ \cite{Gukov:2004id}, thereby realizing the holographic duality prior to the average. It was moreover conjectured in \cite{Gukov:2004id} that the Maxwell-Chern-Simons theory describes the long-distance limit of string theory on AdS$_3\times K_7$, with $K_7$ a compact seven-manifold. Our discussion is therefore relevant not only for three-dimensional gravity but also for string theory \cite{Ashwinkumar:2023jtz}.

\section{Emergent Global Symmetries}\label{sec:emergent}

We have seen that the holographic dual of the ensemble average contains non-trivial anyons. Let us next discuss their global symmetries, following \cite{Ashwinkumar:2023jtz}.

\subsection{Symmetries of Anyons}\label{subsec:anyon_sym}

Abelian Chern-Simons theories have global one-form symmetries, generated by the Abelian anyons themselves, and zero-form symmetries arising from automorphisms of the anyon data \cite{Barkeshli:2014cna,Delmastro:2019vnj}. The one-form symmetry group is therefore identical to the discriminant group $\calD$ (cf.\ \cite{Hsin:2018vcg}). The zero-form symmetry group consists of permutations of anyons which preserve the anyon data $(\calD,\uptheta)$:
\begin{align}
  \Aut(\calD,\uptheta) := \{ g \in \Aut(\calD) \,|\, \uptheta(g\cdot\alpha) = \uptheta(\alpha) \;\; \forall \alpha\in \calD\} \;.
  \label{eq:Aut}
\end{align}
Since the anyon data determine the modular $S$- and $T$-matrices \eqref{eq:ST}, and hence the Gauss sums $\gamma_{Q,\alpha}(c,d)$, we find that the Eisenstein series are invariant under the zero-form symmetry, namely the Eisenstein series for $\alpha$ and $g\cdot\alpha$ coincide:
\begin{align}
  \langle \vartheta_{Q,g\cdot\alpha} \rangle = E_{Q,g\cdot\alpha}(\tau,\bar{\tau}) = E_{Q,\alpha}(\tau,\bar{\tau}) = \langle \vartheta_{Q,\alpha} \rangle \;, \qquad g\in \Aut(\calD,\uptheta) \;.
  \label{eq:sym_after}
\end{align}
This is the expected torus-level consequence of the zero-form symmetries of the bulk TQFT. Equality of these partition-function components alone does not establish a symmetry of every observable in a complete boundary dual.

\subsection{Emergence after Averaging}\label{subsec:emergence}

The question we now address is how these symmetries arise in the process of the ensemble average. To answer this question, let us examine the action of the zero-form symmetry $\alpha\to g\cdot\alpha$ on the partition functions \emph{before} the average.

The zero-form symmetries can be classified as either classical or quantum. The classical symmetries arise from the symmetries of the Lagrangian \eqref{eq:S_CS}: each $\Sigma\in O_Q(p,q;\bbZ)$ in \eqref{eq:T-duality} leaves the Lagrangian invariant under the redefinition $A_i\to\Sigma_{ij}A_j$, acts on $\calD=\Lambda^*/\Lambda$, and preserves the spin \eqref{eq:Aut}. The group of classical symmetries is therefore the image of the natural homomorphism $O_Q(p,q;\bbZ)\to\Aut(\calD,\uptheta)$. This homomorphism in general has a kernel, consisting of the elements acting trivially on $\calD$; for example, for self-dual $Q$ the discriminant group is trivial and so is the image. The group $O_Q(p,q;\bbZ)$ is also the T-duality group of the boundary theory. Quantum symmetries, when present, are not manifest at the level of the Lagrangian; a given lattice need not have any such additional symmetry.

For a quantum symmetry, the symmetry action changes the functional form of the theta function: we in general have
\begin{align}
  \vartheta_{Q,g\cdot\alpha}(\tau,\bar{\tau};m) \neq \vartheta_{Q,\alpha}(\tau,\bar{\tau};m) \;,
  \label{eq:not_sym}
\end{align}
and $g$ is not a symmetry before the ensemble average. The symmetry is therefore truly emergent, and appears only after the ensemble average as in \eqref{eq:sym_after}.

For a classical symmetry $g\in\Aut(\calD,\uptheta)$ in this image, represented by $\Sigma\in O_Q(p,q;\bbZ)$ that moves the Hamiltonian, the inequality \eqref{eq:not_sym} can hold at a generic point of moduli space, although special fixed points may have an enhanced symmetry. There are also exceptions at every point: charge conjugation $\Sigma=-1$ preserves $H$ and gives $\vartheta_{Q,-\alpha}=\vartheta_{Q,\alpha}$ before averaging. In general we have the covariance relation
\begin{align}
  \vartheta_{Q,\Sigma\cdot\alpha}(\tau,\bar{\tau};\Sigma\cdot m) = \vartheta_{Q,\alpha}(\tau,\bar{\tau};m) \;,
  \label{eq:covariance}
\end{align}
where the action $\Sigma\cdot m$ is defined on the covering space of Hamiltonians by $H\to (\Sigma^{-1})^T H \Sigma^{-1}$. This is the covariance of the labeled family of partition functions before taking the T-duality quotient: $\Sigma$ acts both on the anyon label and on a representative of the moduli. After averaging over the appropriate labeled moduli spaces, the resulting relation acts only on the anyon label.

\subsection{Duality Origami}\label{subsec:origami}

The discussion above can be formulated more generally \cite{Ashwinkumar:2023jtz}. Consider an ensemble average of an observable $\calO_{\alpha}(m)$ of a theory $\mathcal{T}(m)$ (CFT or otherwise) over a moduli space $m\in\calM$. In general the integration domain $\calM_{\alpha}\subset\calM$ can depend on the label $\alpha$, as in the case of $\calM_{Q,\alpha}$ in \eqref{eq:average}. Suppose that there exists a group $G$ acting both on the labels $\alpha$ and on the moduli space $\calM$ such that
\begin{enumerate}
  \item $G$ preserves the measure, $[d(g\cdot m)]=[dm]$ for $g\in G$, and maps the integration domains into one another, $g\cdot\calM_{\alpha}=\calM_{g\cdot\alpha}$ (so that in particular $\Vol(\calM_{\alpha})=\Vol(\calM_{g\cdot\alpha})$), and
  \item $G$ acts covariantly on the observable: $\calO_{g\cdot\alpha}(g\cdot m) = \calO_{\alpha}(m)$.
\end{enumerate}
We call such a $G$ an \emph{ensemble symmetry}. Note that an ensemble symmetry is not a symmetry of a theory in the standard sense: an element $g\in G$ maps a point $m$, together with the labels $\alpha$ of the sectors, to a different point $g\cdot m$ with relabeled sectors $g\cdot\alpha$. For T-duality the two describe the same CFT with differently labeled sectors; they are, however, different points of the covering space of the moduli space (before taking the quotient by the T-duality group), and $g$ is not a symmetry of the theory at a fixed $m$.

Once we have an ensemble symmetry, it follows that
\begin{align}
  \begin{split}
    \langle \calO_{\alpha} \rangle
    &= \frac{1}{\Vol(\calM_{\alpha})}\int_{\calM_{\alpha}} [dm]\, \calO_{\alpha}(m)
    = \frac{1}{\Vol(\calM_{\alpha})}\int_{\calM_{\alpha}} [dm]\, \calO_{g\cdot\alpha}(g\cdot m) \\
    &= \frac{1}{\Vol(\calM_{\alpha})}\int_{\calM_{g\cdot\alpha}} [d(g^{-1}\cdot m)]\, \calO_{g\cdot\alpha}(m)
    = \frac{1}{\Vol(\calM_{g\cdot\alpha})}\int_{\calM_{g\cdot\alpha}} [dm]\, \calO_{g\cdot\alpha}(m)
    = \langle \calO_{g\cdot\alpha} \rangle \;,
  \end{split}
\end{align}
where we used the covariance of $\calO$, the change of variables together with $g\cdot\calM_{\alpha}=\calM_{g\cdot\alpha}$, and the invariance of the measure in turn. The averaged observable is invariant. Interpreting this as a symmetry of a complete averaged theory requires the corresponding covariance for its full set of observables.

We can formulate this as a general lesson in ensemble averages:
\begin{tcolorbox}
  A symmetry connecting different points of an ensemble (e.g.\ T-duality) can become a symmetry of ensemble-averaged observables.
\end{tcolorbox}
\noindent In our example, the T-duality action on the labeled family is ``folded'' into an invariance of the averaged quantities. We call this process \emph{duality origami}. Its relation to the holographic argument against global symmetries \cite{Harlow:2018jwu,Harlow:2018tng} depends on how the complete bulk dual is defined. As in the case of the orbifolds \cite{Ashwinkumar:2023eit}, one can moreover gauge part of the ensemble symmetries.

Let us close this section with a remark. An emergent symmetry after the ensemble average can be detected (or not) depending on the choice of observables. For example, even when the ensemble average of a charged observable vanishes, $\langle \calO \rangle = 0$, its averaged squared magnitude can be non-vanishing, $\langle |\calO|^2 \rangle \neq 0$. This is reminiscent of the Edwards--Anderson order parameter in spin glasses \cite{Edwards:1975aas}, where the average of the squared local magnetization detects the order while the average magnetization vanishes. We will come back to this point in \cref{sec:SWSSB}.

\section{Back to the Swampland?}\label{sec:swampland}

Ensemble-averaged theories circumvent several Swampland conjectures, and there have been arguments that ensemble averages are excluded by the Swampland program, at least in ten-dimensional string theory \cite{McNamara:2020uza}. In this section we discuss how the emergent symmetries discussed above can be reconciled with the Swampland program \cite{Ashwinkumar:2023jtz}.

\subsection{Global Symmetries at Infinite Distance}\label{subsec:distance}

In string theory, there is an alternative mechanism for obtaining global symmetries: going to the infinite distance limit of the moduli space. Indeed, the Swampland distance conjecture \cite{Ooguri:2006in} states that there is a tower of states becoming light in the infinite distance limit, where we often encounter an emergence of global symmetries \cite{Grimm:2018ohb,Heidenreich:2018kpg}.

In our case, the natural limit is the limit of the modulus $\tau$ of the spacetime torus: the cusps $\tau\to i\infty$ and its $\PSL(2,\bbZ)$ images $\tau\to a/c\in\mathbb{Q}$, which are at infinite distance in the upper half plane. Here $\tau$ is the complex structure of the torus boundary of the AdS$_3$ part of the AdS$_3\times K_7$ geometry. Each cusp corresponds to a geometry $M_{(c,d)}$ in the bulk, and in the limit to the cusp one of the cycles of the geometry shrinks to zero size, so that the three-dimensional gravity reduces to two-dimensional gravity. This infinite distance limit is different from the usual Swampland distance conjecture, which deals with displacements of the compactification moduli, and is more akin to the generalized distance conjecture of \cite{Lust:2019zwm}.

As we have seen in \eqref{eq:cusp}, the leading behavior of the theta function at the cusp is given by the lens space partition function $\gamma_{Q,\alpha}(c,d)$, which is moduli independent. This is precisely the quantity which encapsulates the global symmetries of the anyons after the ensemble average. On the other hand, the Eisenstein series \eqref{eq:Eisenstein} obtained by the ensemble average is built from exactly the same building blocks $\gamma_{Q,\alpha}(c,d)$. The relations among these quantities are schematically summarized as
\[
  \begin{tikzpicture}[baseline=(current bounding box.center)]
    \node (theta) at (0,1.8) {$\vartheta_{Q,\alpha}(\tau,\bar{\tau};m)$};
    \node (gamma) at (7,1.8) {$\gamma_{Q,\alpha}(c,d)$};
    \node (E) at (0,0) {$\langle\vartheta_{Q,\alpha}\rangle = E_{Q,\alpha}(\tau,\bar{\tau})$};
    \draw[->] (theta) -- node[above]{\small asymptotics at cusps ($\infty$-distance)} (gamma);
    \draw[->] (theta) -- node[left]{\small ensemble average} (E);
    \draw[->] (gamma) -- node[below right]{\small building block} (E);
  \end{tikzpicture}
\]
Namely, the emergence of global symmetries in ensemble averages and that in the infinite distance limit lead to exactly the same expression, the lens space partition function. This is reminiscent of the results of \cite{Collier:2022emf}, where the ensemble average over the conformal manifold of $\mathcal{N}=4$ super Yang-Mills theory is related to the large $N$, large 't Hooft coupling limit. These results hint towards deeper connections between ensemble averages in holography, the Swampland program, and string theory.

\subsection{Fluctuations around Ensemble Averages}\label{subsec:fluctuation}

If we take seriously the correspondence between ensemble averages and the distance conjecture, a natural question is how the emergent global symmetries are broken in honest theories of quantum gravity. In the case of the distance conjecture \cite{Ooguri:2006in}, this is achieved by staying at a finite distance in the moduli space. In analogy with this, one can re-introduce the moduli dependence in the ensemble average. For this purpose we should take the viewpoint that ensemble-averaged holographic correspondences are coarse-grained descriptions of microscopic dualities, and the moduli dependence represents fluctuations from the ensemble average.

Remarkably, the relevant mathematics is already known as the Roelcke--Selberg spectral decomposition \cite{Roelcke:1966,Terras:1985}, which was applied to the partition functions of Narain CFTs in \cite{Benjamin:2021ygh}. For a square-integrable modular form $f(\tau,\bar{\tau})$ for a congruence subgroup, the decomposition reads
\begin{align}
  f(\tau,\bar{\tau}) = \sum_{i} \langle f, u_i\rangle u_i(\tau) + \sum_{j} \langle f, v_j\rangle v_j(\tau) + \sum_{k} \frac{1}{4\pi} \int_{-\infty}^{\infty} dt\, \left\langle f, E_{\mathfrak{a}_k}\!\left(\cdot, \tfrac{1}{2}+it\right) \right\rangle E_{\mathfrak{a}_k}\!\left(\tau, \tfrac{1}{2}+it\right) \;,
\end{align}
where $\{u_i\}$ is an orthonormal basis of cusp forms, $\{v_j\}$ are residues of the Eisenstein series, $E_{\mathfrak{a}_k}$ are Eisenstein series labeled by the cusps $\mathfrak{a}_k$, and $\langle -,-\rangle$ is the Petersson inner product. When we apply this decomposition to the combination
\begin{align}
  f(\tau,\bar{\tau}) = \tau_2^{\frac{p+q}{4}} \left( \vartheta_{Q,\alpha}(\tau,\bar{\tau};m) - E_{Q,\alpha}(\tau,\bar{\tau}) \right) \;,
\end{align}
which measures the deviation from the ensemble average, all the terms in the decomposition are moduli dependent, and trigger the breaking of the emergent global symmetries. This suggests that the counterpart of the ``tower of states'' in the Swampland distance conjecture should be included in the moduli-dependent terms of the spectral decomposition.

\section{Strong-to-Weak Spontaneous Symmetry Breaking}\label{sec:SWSSB}

So far we discussed ensembles of CFTs parametrized by a moduli space. One may ask if the lessons learned above apply to more general ensembles. One natural candidate is the classical probability distribution in quantum theory, namely the density matrix
\begin{align}
  \hat{\rho} = \sum_i p_i\, |\psi_i\rangle \langle \psi_i| \;.
\end{align}
In this section we discuss the emergence of symmetries for density matrices, following \cite{Kawamoto:2026jdz}.

\subsection{Strong and Weak Symmetries}\label{subsec:strong_weak}

A symmetry of a pure state $|\psi\rangle$ is described by a unitary operator $\hat{U}$ acting as $\hat{U}|\psi\rangle \propto |\psi\rangle$, and this preserves the density matrix $\hat{\rho}=|\psi\rangle\langle\psi|$ as $\hat{U}\hat{\rho}\, \hat{U}^{\dagger}=\hat{\rho}$. For mixed states, we need to distinguish two different notions of symmetries \cite{Buca:2012zz,Albert:2014yej,deGroot:2021zot}:
\begin{itemize}
  \item \emph{weak symmetry}: $\hat{U}\hat{\rho}\,\hat{U}^{\dagger} = \hat{\rho}$,
  \item \emph{strong symmetry}: $\hat{U}\hat{\rho} = e^{i\theta} \hat{\rho}$.
\end{itemize}
A strong symmetry is automatically a weak symmetry, but not conversely.

As an example, consider $G=U(1)$ and denote by $|\psi_{\mathsf{Q}}\rangle$ a state with charge $\mathsf{Q}$. A mixture of different charge sectors
\begin{align}
  \hat{\rho} = \sum_{\mathsf{Q}} c_{\mathsf{Q}}\, |\psi_{\mathsf{Q}}\rangle \langle \psi_{\mathsf{Q}}|
  \label{eq:rho_weak}
\end{align}
is weakly symmetric but not strongly symmetric. By contrast, a density matrix supported in a fixed charge sector, e.g.\
\begin{align}
  \hat{\rho} = \hat{\rho}_{\mathsf{Q}} = \sum_i p_i\, |\psi_{\mathsf{Q},i}\rangle \langle \psi_{\mathsf{Q},i}|
\end{align}
for a specific $\mathsf{Q}$, where $|\psi_{\mathsf{Q},i}\rangle$ are states with charge $\mathsf{Q}$, is strongly symmetric: $e^{i\theta\hat{\mathsf{Q}}}\hat{\rho}_{\mathsf{Q}} = e^{i\theta \mathsf{Q}}\hat{\rho}_{\mathsf{Q}}$, where $\hat{\mathsf{Q}}$ is the $U(1)$ charge operator.

Earlier work studied symmetry breaking in open systems \cite{Minami:2015uzo,Hongo:2018ant,Hidaka:2019irz}. For mixed states, the distinction between strong and weak symmetries allows for a pattern with no closed-system analog: strong-to-weak spontaneous symmetry breaking (SWSSB), where a strong symmetry is spontaneously broken down to a weak symmetry \cite{Lee:2023fsk,Ma:2023rji,Lessa:2024gcw,Sala:2024ply}. This should be contrasted with conventional (strong) spontaneous symmetry breaking, where the symmetry is completely broken:
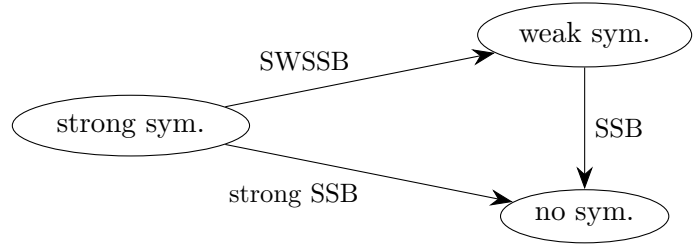
\[
  \begin{tikzpicture}[baseline=(current bounding box.center),>={Stealth[length=3mm,width=2.2mm]}]
    \node[draw,ellipse] (strong) at (0,0) {strong sym.};
    \node[draw,ellipse] (weak) at (6,1.2) {weak sym.};
    \node[draw,ellipse] (none) at (6,-1.2) {no sym.};
    \draw[->] (strong) -- node[above left]{\small SWSSB} (weak);
    \draw[->] (strong) -- node[below left]{\small strong SSB} (none);
    \draw[->] (weak) -- node[right]{\small SSB} (none);
  \end{tikzpicture}
\]

These notions are best understood in the doubled Hilbert space. Choose an orthonormal basis $\{|n\rangle\}$ of $\calH$, let $\calH_-$ denote its conjugate space, and define $|I\rangle\!\rangle:=\sum_n|n\rangle_+\otimes|n^*\rangle_-$. By the Choi--Jamio{\l}kowski map, a density matrix becomes a state in $\calH_+\otimes \calH_-$:
\begin{align}
  \hat{\rho} = \sum_i p_i\, |\psi_i\rangle\langle\psi_i| \quad \longrightarrow \quad |\rho\rangle\!\rangle := \sum_i p_i\, |\psi_i\rangle_+ \otimes |\psi_i^*\rangle_- \;.
\end{align}
This doubling is natural from the point of view of the Schwinger--Keldysh path integral \cite{Schwinger:1960qe,Keldysh:1964ud}:
\begin{align}
  e^{W[J_+,J_-]} = \int_{\rho} D\phi_+ D\phi_-\, e^{iS_+[\phi_+,J_+]-iS_-[\phi_-,J_-]}
  = \langle\!\langle I| \, \hat{U}(J_+) \otimes \hat{U}(J_-)^* \, |\rho\rangle\!\rangle \;,
\end{align}
where $\hat{U}(J_{\pm})$ are the time-evolution operators in the presence of sources $J_{\pm}$, and the information on the initial state is encoded in the junction conditions interpolating between $\phi_+$ and $\phi_-$. With these conventions, $\langle\!\langle I|A\otimes B^*|\rho\rangle\!\rangle=\Tr[A\hat\rho B^{\dagger}]$. The Schwinger--Keldysh action enjoys the doubled symmetry $G_+\times G_-$, which is broken by the choice of the state $\rho$. A strong symmetry is the symmetry under $G_+\times G_-$, while a weak symmetry is the symmetry under the diagonal subgroup.

\subsection{SWSSB as Emergent Symmetry in Ensemble Averages}\label{subsec:SWSSB_emergent}

The formulation above makes it clear that SWSSB is closely related to the emergence of symmetries in ensemble averages. Let us start with a pure state $|\psi\rangle = \sum_{\mathsf{Q}} a_{\mathsf{Q}} |\psi_{\mathsf{Q}}\rangle$, which is a superposition of states with different charges, and define $|\psi\rangle\!\rangle:=|\psi\rangle_+\otimes|\psi^*\rangle_-$. This state has no symmetry, not even a weak symmetry, since the corresponding density matrix contains off-diagonal terms between different charge sectors:
\begin{align}
  \langle\!\langle I| \, \hat{U}(J_+) \otimes \hat{U}(J_-)^* \, |\psi\rangle\!\rangle
  = \sum_{\mathsf{Q},\mathsf{Q}'} a_{\mathsf{Q}} a^*_{\mathsf{Q}'}\, \langle \psi_{\mathsf{Q}'}| \hat{U}(J_-)^{\dagger}\hat{U}(J_+) |\psi_{\mathsf{Q}}\rangle \;.
\end{align}
The terms with $\mathsf{Q}\neq \mathsf{Q}'$ are not invariant under the diagonal $G$. Suppose, however, that we average over the phases of the coefficients $a_{\mathsf{Q}}$. This is the same as the group average (known also as twirling)
\begin{align}
  \hat{\rho} = \int dg\, \hat{U}_g |\psi\rangle\langle\psi| \hat{U}_g^{\dagger} \;,
  \label{eq:twirling}
\end{align}
with $dg$ the normalized Haar measure, and the off-diagonal terms are washed out. We then obtain the density matrix \eqref{eq:rho_weak} with $c_{\mathsf{Q}}=|a_{\mathsf{Q}}|^2$, which has a weak symmetry. The weak symmetry, therefore, emerges as a result of the ensemble average. In the Schwinger--Keldysh path integral, the generating functional for the twirled state \eqref{eq:twirling} reads
\begin{align}
  \begin{split}
    e^{W[J_+,J_-]} &= \Tr\left[ \hat{U}(J_+)\, \hat{\rho}\, \hat{U}(J_-)^{\dagger} \right]
    = \int dg\, \sum_{n} Z_g^{J_+}(n,\psi)\, Z_g^{J_-}(n,\psi)^* \;, \\
    Z_g^{J}(n,\psi) &:= \langle n| \hat{U}(J)\, \hat{U}_g |\psi\rangle \;,
  \end{split}
\end{align}
where $\{|n\rangle\}$ is the orthonormal basis used in the definition of $|I\rangle\!\rangle$. Note that we need different sources $J_+\neq J_-$ on the two branches: for $J_+=J_-$ the generating functional reduces to $\Tr\,\hat{\rho}=1$. Each path integral $Z_g^{J}(n,\psi)$ does not possess the symmetry, since the boundary condition associated with $|\psi\rangle$ explicitly breaks the symmetry. The weak symmetry emerges only after the ensemble average over $g$. This is precisely parallel to the emergence of symmetries in the ensemble averages discussed in \cref{sec:emergent}.

More generally, consider a mixed state obtained by a deformation which keeps the weak symmetry but not the strong symmetry:
\begin{align}
  |\rho\rangle\!\rangle = \exp\left( \sum_q \lambda_q \int_M \calO^{[q]}_+ \calO^{[-q]}_- \right) |\psi_+\rangle|\varphi_-\rangle \;,
\end{align}
where $M$ is the spacetime region over which the deformation acts, $|\psi_+\rangle$ and $|\varphi_-\rangle$ are invariant under the symmetry, and $\calO^{[\pm q]}_{\pm}$ are operators with charges $\pm q$ in the two copies. For a density operator the undeformed ket and bra must be paired appropriately, for example $|\varphi_-\rangle=|\psi_+^*\rangle$; further conditions on the deformation ensure Hermiticity and positivity. By the Hubbard--Stratonovich transformation, the Schwinger--Keldysh path integral can be rewritten as \cite{Kawamoto:2026jdz}
\begin{align}
  e^{W[J_+,J_-]} = \int D\chi_+ D\chi_-\, P(\chi_+,\chi_-) \prod_{s=\pm} Z_{s}[J_{s};\chi_{s}] \;,
\end{align}
where $P(\chi_+,\chi_-)$ is a quasi-probability distribution for the auxiliary fields $\chi^{[q]}_{\pm}$, and the factor $Z_{\pm}[J_{\pm};\chi_{\pm}]$ is the path integral in the presence of the source term $\sum_q \chi^{[q]}_{\pm}\calO^{[\pm q]}_{\pm}$. Each $Z_{\pm}[J_{\pm};\chi_{\pm}]$ does not possess the symmetry, which is explicitly broken by the deformation. The diagonal symmetry is, however, restored by integrating over the auxiliary fields with the weight $P(\chi_+,\chi_-)$: the symmetry emerges through the ensemble average.
These constructions give an ensemble interpretation of SWSSB: a weak symmetry emerges when symmetry-breaking ingredients are averaged. 

\subsection{Diagnosing SWSSB}\label{subsec:diagnose}

Let us next discuss how to diagnose SWSSB. For concreteness, consider $G=U(1)$ and a charged local operator $\hat{\calO}(x)$ with charge $q\neq 0$. The R\'enyi-2 correlator is defined by \cite{Lee:2023fsk,Lessa:2024gcw}
\begin{align}
  R^{(2)}(x,y) := \frac{\Tr\left[ \hat{\calO}(x)\hat{\calO}(y)^{\dagger}\, \hat{\rho}\, \hat{\calO}(y)\hat{\calO}(x)^{\dagger}\, \hat{\rho} \right]}{\Tr\left[\hat{\rho}^2\right]} \;,
\end{align}
Writing $X_{xy}:=\hat{\calO}(x)\hat{\calO}(y)^{\dagger}$, the Choi convention used above gives
\begin{align}
  R^{(2)}(x,y) = \frac{\langle\!\langle\rho|\, X_{xy}\otimes X_{xy}^{*}\, |\rho\rangle\!\rangle}{\langle\!\langle\rho|\rho\rangle\!\rangle} \;,
\end{align}
where $*$ denotes complex conjugation in the basis $\{|n\rangle\}$.
A mixed state $\hat{\rho}$ with a strong $U(1)$ symmetry has SWSSB in the R\'enyi-2 sense if there exists a charged local operator such that (a) the R\'enyi-2 correlator has long-range order\footnote{One can also use the fidelity correlator or the Wightman correlator (the R\'enyi-1 correlator) as diagnostics of SWSSB \cite{Lessa:2024gcw}.}\begin{align}
  \lim_{|x-y|\to\infty} R^{(2)}(x,y) = O(1) \;,
  \label{eq:SWSSB_a}
\end{align}
while (b) the conventional correlation functions show no long-range order for any charged operators
\begin{align}
  \lim_{|x-y|\to\infty} \Tr\left[ \hat{\rho}\, \hat{\calO}(x)\hat{\calO}(y)^{\dagger} \right] = 0 \;,
  \label{eq:SWSSB_b}
\end{align}
so that the weak symmetry is not spontaneously broken. Note that the R\'enyi-2 correlator involves two copies of the density matrix, and is quadratic in $\hat{\rho}$. The situation is analogous to the remark at the end of \cref{sec:emergent}: the ordinary correlator \eqref{eq:SWSSB_b}, linear in $\hat{\rho}$, vanishes, while the correlator \eqref{eq:SWSSB_a}, quadratic in $\hat{\rho}$, detects the order, just as for the Edwards--Anderson order parameter in spin glasses \cite{Edwards:1975aas} (see also \cite{Lessa:2024gcw}).

\subsection{How General is SWSSB?}\label{subsec:general}

We now focus on SWSSB at finite temperature. Following \cite{Lessa:2024gcw,Sala:2024ply}, consider the density matrix
\begin{align}
  \hat{\rho}_{\mathsf{Q},\beta} := \frac{\hat{\Pi}_{\mathsf{Q}}\, e^{-\beta\hat{H}}}{\Tr\left[\hat{\Pi}_{\mathsf{Q}}\, e^{-\beta\hat{H}}\right]} \;,
  \label{eq:rho_Q_beta}
\end{align}
where $\hat{H}$ is a $U(1)$-symmetric Hamiltonian, $\beta\in(0,\infty)$ is the inverse temperature, and $\hat{\Pi}_{\mathsf{Q}}$ is the projector onto the sector with a fixed charge $\hat{\mathsf{Q}}=\mathsf{Q}\in\bbZ$. This state is strongly symmetric, in contrast with the grand canonical ensemble $e^{-\beta(\hat{H}-\mu\hat{\mathsf{Q}})}/\Tr[e^{-\beta(\hat{H}-\mu\hat{\mathsf{Q}})}]$, which has only a weak symmetry. For this choice of the density matrix it has been conjectured that SWSSB is ubiquitous in the thermodynamic limit \cite{Lessa:2024gcw,Sala:2024ply}:
\begin{tcolorbox}
  \textbf{Lattice SWSSB Conjecture} \cite{Lessa:2024gcw,Sala:2024ply,Kawamoto:2026jdz}: Suppose that we have a local $U(1)$-symmetric Hamiltonian $\hat{H}$, an inverse temperature $0<\beta<\infty$, and the density matrix $\hat{\rho}_{\mathsf{Q},\beta}$ \eqref{eq:rho_Q_beta} with a strong $U(1)$ symmetry. If $\hat{\rho}_{\mathsf{Q},\beta}$ has no SSB, then it must have SWSSB in the R\'enyi-2 sense used here.
\end{tcolorbox}

Let us sketch the reasoning behind the conjecture \cite{Kawamoto:2026jdz}. Since the projector $\hat{\Pi}_{\mathsf{Q}}$ commutes with the neutral operator $\hat{\calO}(x)\hat{\calO}(y)^{\dagger}$, and since $(\hat{\rho}_{\mathsf{Q},\beta})^2\propto \hat{\rho}_{\mathsf{Q},2\beta}$, we can rewrite the R\'enyi-2 correlator as a thermal four-point function at the doubled inverse temperature:
\begin{align}
  R^{(2)}(x,y) = \left\langle \hat{\calO}(x;-i\beta)\hat{\calO}^{\dagger}(y;-i\beta)\, \hat{\calO}(y)\hat{\calO}^{\dagger}(x) \right\rangle_{\mathsf{Q},2\beta} \;,
  \label{eq:R2_four_point}
\end{align}
Here $\langle\cdots\rangle_{\mathsf{Q},2\beta}:=\Tr[\hat{\rho}_{\mathsf{Q},2\beta}\cdots]$ and $\hat{\calO}^{\dagger}(x;t):=e^{i\hat Ht}\hat{\calO}(x)^{\dagger}e^{-i\hat Ht}$; at complex $t$ this differs from $\hat{\calO}(x;t)^{\dagger}$. The fixed-charge state has $\langle\hat{\calO}(x)\rangle_{\mathsf{Q},2\beta}=0$. If the weak symmetry remains unbroken and the state clusters at inverse temperature $2\beta$, the separated charged two-point function vanishes and cluster decomposition gives
\begin{align}
  R^{(2)}(x,y) \approx \left\langle \hat{\calO}(x;-i\beta)\hat{\calO}^{\dagger}(x) \right\rangle_{\mathsf{Q},2\beta} \left\langle \hat{\calO}^{\dagger}(y;-i\beta)\hat{\calO}(y) \right\rangle_{\mathsf{Q},2\beta} \qquad (|x-y|\to\infty)\;.
\end{align}
Assuming local equivalence of the fixed-charge and grand canonical ensembles for these two-point functions, together with reflection positivity, we have
\begin{align}
  \left\langle \hat{\calO}(x;-i\beta)\hat{\calO}^{\dagger}(x) \right\rangle_{\mathsf{Q},2\beta}
  \approx \left\langle \hat{\calO}(x;-i\beta)\hat{\calO}^{\dagger}(x) \right\rangle_{\mu,2\beta}
  = \left\langle \hat{\calO}\left(x;-\tfrac{i\beta}{2}\right)\hat{\calO}^{\dagger}\left(x;\tfrac{i\beta}{2}\right) \right\rangle_{\mu,2\beta} > 0 \;,
\end{align}
where the chemical potential $\mu$ is chosen to reproduce the charge density $\mathsf{Q}/V$ in the thermodynamic limit, with $V$ the spatial volume of the system. Under these assumptions, and provided the positive local factor stays bounded away from zero as the volume grows, we obtain SWSSB for the R\'enyi-2 correlator. The conjecture assumes no ordinary SSB at inverse temperature $\beta$, whereas this argument also needs clustering at $2\beta$. Local ensemble equivalence and a volume-uniform lower bound must likewise be checked for a given system \cite{Kawamoto:2026jdz}. We next turn to holographic theories.
 
\subsection{SWSSB in Holography}\label{subsec:holography}

Let us consider the bottom-up AdS$_3$/CFT$_2$ correspondence (see e.g.\ \cite{Kraus:2006wn}) with a $U(1)$ global symmetry on the boundary, which is a gauge symmetry in the bulk. Following \cite{Kawamoto:2026jdz}, we take Einstein gravity coupled to two Abelian Chern-Simons gauge fields $A$ and $\bar A$ of opposite levels, together with a charged complex bulk scalar $\mathsf{\Psi}$ that serves as a probe dual to $\hat{\calO}$. Since the Chern-Simons sector is topological, the gravitational background can be treated separately at probe order. The dominant saddle is either the BTZ black hole or thermal AdS$_3$, depending on the temperature \cite{Hawking:1982dh,Witten:1998zw}; in the infinite-volume limit, which we need to discuss long-range order, the dominant contribution is given by the BTZ black hole.

The R\'enyi-2 correlator \eqref{eq:R2_four_point} is a four-point function on the thermal circle of length $2\beta$, where the pairs of operators are inserted at the imaginary times $0$ and $\beta$. In the bulk, the two insertion points are connected through the Euclidean geometry. In the large $N$ limit, where $N$ schematically denotes the number of degrees of freedom of the boundary CFT (so that the central charge scales as $\mathsf{c}\sim N$ and the bulk is weakly coupled), charge conservation leaves two leading factorization channels. Schematically, when the charge projection is controlled by a common chemical-potential saddle, they are
\begin{align}
  \begin{split}
    R^{(2)}(x,y) \approx {}&
    \big\langle\hat{\calO}(x;-i\beta)\hat{\calO}^{\dagger}(x)\big\rangle_{\mathsf{Q},2\beta}
    \big\langle\hat{\calO}^{\dagger}(y;-i\beta)\hat{\calO}(y)\big\rangle_{\mathsf{Q},2\beta} \\
    &+\big\langle\hat{\calO}(x;-i\beta)\hat{\calO}^{\dagger}(y;-i\beta)\big\rangle_{\mathsf{Q},2\beta}
    \big\langle\hat{\calO}(y)\hat{\calO}^{\dagger}(x)\big\rangle_{\mathsf{Q},2\beta} \;,
  \end{split}
\end{align}
In the limit $|x-y|\to\infty$ only the first term, the product of the two-point functions across the two halves of the thermal circle, survives. This is a holographic counterpart of cluster decomposition. More precisely, the projectors $\hat{\Pi}_{\mathsf{Q}}$ in the two factors originate from a single integral over the chemical potential $\mu$ (see \eqref{eq:projector} below). The large $N$ factorization should therefore be applied at fixed $\mu$, and the integral over $\mu$ is performed afterwards.

It remains to show that the resulting left-right contribution is non-vanishing. This requires the left-right two-point function at fixed $\mu$, which we denote by $G_{\mu}(\beta,0)$: the two-point function of the charged operator between the imaginary times $\beta$ and $0$ at the same spatial point, on the thermal circle of circumference $2\beta$ with the chemical potential $\mu$ turned on. Its computation is essentially a standard holographic computation, except that we need a bulk implementation of the projector $\hat{\Pi}_{\mathsf{Q}}$. In CFTs with a $U(1)$ symmetry, we can write the projector as an integral over the imaginary chemical potential:
\begin{align}
  \hat{\Pi}_{\mathsf{Q}} = \int_0^{2\pi} \frac{d\mu}{2\pi}\, e^{i\mu(\hat{\mathsf{Q}}-\mathsf{Q})} \;.
  \label{eq:projector}
\end{align}
In the bulk, each operator $e^{i\mu\hat{\mathsf{Q}}}$ is represented by a Wilson line defect \cite{Zhao:2020qmn}; in the BTZ geometry it is inserted along the horizon cycle $C$, and it induces a holonomy of the gauge fields around the Euclidean time circle linking $C$. For the thermal circle of circumference $2\beta$ used in \eqref{eq:R2_four_point}, the charged scalar then obeys $\mathsf{\Psi}(t_E+2\beta)=e^{+i\mu q}\mathsf{\Psi}(t_E)$. It is convenient to remove the gauge fields from the wave equation by the gauge rotation $\mathsf{\Psi}=\exp\bigl(iq\int A+i\bar{q}\int\bar{A}\bigr)\mathsf{\Phi}$, with the line integrals taken along the radial direction from the boundary \cite{Kawamoto:2026jdz}. The rotated field $\mathsf{\Phi}$ satisfies the free wave equation and the twisted boundary condition $\mathsf{\Phi}(t_E+2\beta)=e^{-i\mu q}\mathsf{\Phi}(t_E)$. The Matsubara frequencies are shifted to
\begin{align} 
  \omega_n = \frac{\pi}{\beta}\left(n+\frac{\mu q}{2\pi}\right) \;, \qquad n\in\bbZ \;.
\end{align}
By solving the wave equation in the BTZ background, the ordinary two-point function is found to have only short-range correlations in the spatial direction, so that there is no conventional SSB. This is as expected: conventional SSB of a continuous symmetry is forbidden at finite temperature in two spacetime dimensions by the Mermin--Wagner--Coleman theorem \cite{Mermin:1966fe,Coleman:1973ci}. The theorem, however, does not apply to the R\'enyi-2 long-range order discussed here. By contrast, after the integral over $\mu$, the R\'enyi-2 correlator in the large-distance limit becomes a weighted average of $|G_{\mu}(\beta,0)|^2$ \cite{Kawamoto:2026jdz}:
\begin{align}
  \lim_{|x-y|\to\infty} R^{(2)}(x,y) = \int_0^{2\pi} d\mu\, w(\mu)\, \bigl|G_{\mu}(\beta,0)\bigr|^2 \;, \qquad
  w(\mu) := \frac{e^{-i\mu\mathsf{Q}}\,\Tr\bigl[e^{i\mu\hat{\mathsf{Q}}}e^{-2\beta\hat{H}}\bigr]}{\int_0^{2\pi} d\mu'\, e^{-i\mu'\mathsf{Q}}\,\Tr\bigl[e^{i\mu'\hat{\mathsf{Q}}}e^{-2\beta\hat{H}}\bigr]} \;.
\end{align}
Note that this is an average of products at fixed $\mu$, rather than a product of the $\mu$-averaged two-point functions. Evaluating this expression, one finds that it is bounded from below by a positive constant independent of the system size \cite{Kawamoto:2026jdz}. This demonstrates SWSSB in this holographic model.

The non-vanishing of the R\'enyi-2 correlator is associated with connectivity of the bulk spacetime, which permits the factorized channel across the two halves of the thermal circle. On the boundary side, the projector $\hat{\Pi}_{\mathsf{Q}}$ acts as a codimension-one defect. In an AdS/BCFT-inspired picture \cite{Takayanagi:2011zk}, one might model a strongly symmetric state by a bulk brane opaque to charged fields and gauge fields, while neutral fields pass through. Such a brane would make the cross-circle charged two-point functions vanish. Their nonzero value in the model instead motivates a connected bulk geometry: a wormhole joining the two sides of the thermofield-double construction \cite{Maldacena:2001kr}, in the spirit of ER=EPR \cite{Maldacena:2013xja}. Local equivalence with the grand canonical ensemble supports this interpretation. In this setup, SWSSB provides evidence for the connected geometry \cite{Kawamoto:2026jdz}.
 
\subsection{Swampland SWSSB Conjecture}\label{subsec:swampland_SWSSB}

Let us finally discuss the interplay between SWSSB and the Swampland program \cite{Kawamoto:2026jdz}. According to the no global symmetry conjecture, even a weak symmetry is forbidden in quantum gravity. As we have seen, however, emergent global symmetries appear in ensemble-averaged theories, and while the necessity of ensemble averages is under debate, they do seem to arise in some simple theories of gravity. One may therefore be tempted to extend the no global symmetry conjecture to more general holographic theories involving ensemble averages. Taking advantage of the connection between emergent symmetries and SWSSB, we propose:
\begin{tcolorbox}
  \textbf{Swampland SWSSB Conjecture} \cite{Kawamoto:2026jdz}: Consider any theory of quantum gravity with a description as weakly-coupled Einstein gravity, with or without ensemble averages in the description. Then any strong global symmetry of a mixed state should be broken at least down to a weak symmetry; namely, the state should exhibit either SWSSB or strong SSB.
\end{tcolorbox}
\noindent It would be interesting to either prove or find a counterexample to this conjecture in string theory compactifications. For $n$ replicas of the doubled construction, one starts with multiple copies of the bulk gauge sector. Wormhole connectivity can identify their symmetries diagonally, while bulk Higgsing can break them. This suggests a possible relation to proposed bounds on massless fields in quantum gravity, but does not by itself establish such a bound \cite{Brennan:2017rbf,Kawamoto:2026jdz}.

\section{Summary and Outlook}\label{sec:summary}

In this article we discussed the fate of global symmetries in quantum gravity from the viewpoint of ensemble averages. In the settings considered here, global symmetries
\begin{itemize}
  \item are ``forbidden'' in quantum gravity,
  \item emerge in ensemble averages, where (T-)duality symmetries relating different theories in the ensemble are folded into global symmetries of a single theory (duality origami),
  \item emerge in the infinite distance limit, and the resulting expressions coincide with those obtained from the ensemble average, and
  \item can arise in an ensemble interpretation of SWSSB of mixed states, for which connected bulk geometries provide a holographic realization.
\end{itemize}

There are many open questions. While our precision case study focused on the generalized Narain theories, it would be interesting to study more general ensembles, e.g.\ ensembles over the moduli spaces of supersymmetric CFTs. This would lead to interesting connections with the geometry of the moduli spaces, e.g.\ of Calabi-Yau manifolds, and with number theory, as we have already seen in the appearance of the Siegel--Weil formula. The moduli-dependent terms in the spectral decomposition describe how the averaged symmetries fail in individual theories. It is also important to clarify the embedding of ensemble averages into string theory, the role of ensemble averages in the factorization puzzle, and whether the proposed Swampland SWSSB Conjecture holds beyond the holographic model discussed here.

Symmetry has been a guiding principle of physics, and we hope that it will continue to guide us in quantum gravity, even in the absence of exact global symmetries.

\section*{Acknowledgements}\addcontentsline{toc}{section}{Acknowledgements}
The author would like to thank Meer Ashwinkumar, Matthew Dodelson, Taishi Kawamoto, Abhiram Kidambi, Jacob M.\ Leedom, and Kenya Tasuki for collaborations on which this article is based, and RIKEN iTHEMS for the invitation to give the lecture (October 24, 2025) from which this article originated.\footnote{The slides for the lecture are available at \url{https://member.ipmu.jp/masahito.yamazaki/files/2025/20251024_RIKEN.pdf}.}
This research was supported in part by the World Premier International Research Center Initiative (WPI), MEXT, Japan; JSPS KAKENHI Grant No.~23K25865; JST, Japan (CREST Grant No.~JPMJCR26XA, Moonshot R\&D Grant No.~JPMJMS256E); and IBM-UTokyo-sponsored research. The author was assisted by Anthropic Claude in writing and editing the manuscript.

\bibliographystyle{JHEP}
\bibliography{refs}

\end{document}